# Statistical Network Topology for Crisis Informetrics

*Liaquat Hossain (Corresponding Author)*
Professor, Information Management
Division of Information and Technology Studies
The University of Hong Kong
lhossain@hku.hk

Honorary Professor, Complex Systems
School of Civil Engineering
Faculty of Engineering and IT
The University of Sydney, Australia
Liaquat.hossain@sydney.edu.au

*Rolf T Wigand*
Maulden-Entergy Chair & Distinguished Professor
Departments of Information Science & Management
UALR, 548 EIT Building
2801 South University Avenue
Little Rock, AR 72204-1099, USA

Shahadat Uddin
Complex Systems
School of Civil Engineering
Faculty of Engineering and IT
The University of Sydney, Australia
Shahadat.uddin@sydney.edu.au

**Crisis informetrics is considered to be a relatively new and emerging area of research, which deals with the application of analytical approaches of network and information science combined with experimental learning approaches of statistical mechanics to explore communication and information flow, robustness as well as tolerance of complex crisis networks under threats. In this paper, we discuss the scale free network property of an organizational communication network and test both traditional (static) and dynamic topology of social networks during organizational crises Both types of topologies exhibit similar characteristics of prominent actors reinforcing the power law distribution nature of scale free networks. There are no significant fluctuations among the actor prominence in daily and aggregated networks. We found that email communication network display a high degree of scale free behavior described by power law.**

A communication network is a pattern of contacts which are created by the flow of data, information, knowledge, etc. among the participating actors (or communicators). Examples of communication networks include personal contact networks, work related contact networks, strategic alliances among various firms, and global networks of organizations. (*1*).

One of the 'self-evident' (*2*) views of organization is described as dynamic systems of adaptation and evolution that contain several parts. These parts also interact with one another and internal and external environments that generate certain outputs. This representation of organizations as dynamic and adaptive system also implies that organizations are 'complex systems'. Complex systems change inputs to outputs in a nonlinear way because their components interact with one another via a web of feedback loops (*3*). Some of the characteristics of the complex systems which could be related to organizational phenomena are: like organizations, complex systems contain large number of interacting agents and associated emerging properties; organizations are complex systems because they are comprised of many individuals, groups and departments that interact with each other and constantly provide feedback; and, finally, like complexity systems, organizations also show emergent properties or behavior which evolves due to the collective behavior of the various interacting agents (*2*, *3*).

In recent years, our understanding of complex networks have changed significantly due to the availability of 'real world' networks (*4*) coupled with the advances in the knowledge base of analytic techniques employed by social network researchers in the area as diverse as Physics and Biology; Mathematics and Sociology; Organizational Science and Psychology. Interestingly, the most commonly cited link between these diverse areas of research is self organising behavior of complex system. Two of the most frequently mentioned properties of real world complex systems are: clustering behavior and the existence of scale free networks (*4*). Research indicates that most networks display a high degree of clustering; and many scientific, technical and organizational networks, ranging from biological networks (*5*) to World Wide Web (*6*) have been found to be scale free. Scale free networks display the characteristics of power law distribution, which states that the probability that a randomly selected node has $k$ links (i.e. degree $k$) follows $P(k) \sim k^{-\gamma}$, where $\gamma$ is the degree exponent (*7*). In this paper, we focus on the scale free network property of the communication network.

We analysed the Enron corpus, email communication log, released by Federal Energy Regulatory Commission (FERC) in May, 2002. Founded in 1985 in Texas, within a decade, Enron became a global player and a symbol of innovative and progressive business conglomerate that also became actively involved in the area of metals, pulps and paper, broadband assets, water plants and financial markets internationally (*8*). In 2000, Enron's annual revenue was $101 billion which made it the seventh largest company in the United States, bigger than IBM or Sony (*9*). However, in 2001, it became slowly evident that with the help of Arthur Andersen (Enron's auditor since 1985), Enron had been grossly overstating its profits and understating debts for the previous five years. On October 16, 2001, Enron disclosed that it had lost $618 million in third quarter earnings. Then, on December 2, 2001, Enron filed for bankruptcy protection in a New York Bankruptcy court. With $62 billion in assets, this was the largest bankruptcy in the history of the US up to that time. By January 2002, Enron stock lost 99% of its value. Stockholders lost tens of billions of dollars and many of the company's 20,000 employees lost their retirement savings pensions and jobs (*8*, *9*, *10*). Following the demise, the FERC publicly released a large set of email messages, the Enron corpus. Details about the dataset can be found in supporting online material. In the area of organizational science and social networking research, the Enron corpus is of great value because it allows academics to conduct research on real-life organization over a number of years.

The aim of this research is to study the scale free network property of the Enron email communication network. We adopt a social network analysis measure of centrality to study the network. Accordingly, we tested both static and dynamic topology of social network analysis during Enron's crisis period that captures the dynamics of Enron's communication network. Starting with the premise that email networks constitute a useful proxy for the underlying communication networks within organizations. With the rapid advancement of information and communication technology, many organizations have been working in virtual environments. Technology has enabled the systematic decentralisation of works. It has also enabled organizational members to work collaboratively even though they are geographically and spatially dispersed. Researchers (*11*, *12*) have argued that computer supported social networks sustain strong, intermediate and weak ties that provide information and social support in both specialized and broadly-based relationships. Others (*13*) argued that email networks provide an inexpensive but powerful alternative to the traditional approach of conducting surveys which is expensive in terms of time and money. Indeed, the exchange of email between individuals in organizations reveals how people interact and facilitates mapping informal networks in a non-intrusive, objective, and quantitative way. Email communication networks are described (*14*) as a tantalizing medium for research, offering a promising resource for tapping into the dynamics of information within organizations and for extracting the hidden patterns of collaboration. Others (*15*) also posited that analysis of email and other interaction logs of organizations will enable researchers to discern the structure of networks and identify core contributors.

Barabási & Albert (*16*) proposed that, independent of the system and the identity of its constituents, the probability $P(k)$ that a vertex in the network interacts with $k$ other vertices decays as a power law, following:

$P(k) \sim k^{-\gamma}$ (1)

They have called this scale-free state, a feature unpredicted by all existing random network models. They propose a model incorporating growth and preferential attachment, two key features of real life networks, and showed that these features are associated with the power-law distribution properties observed in many real networks. Preferential attachment predicts that, in deciding which node to connect to, the new node will prefer the node which is the more connected node. It also implies that highly connected nodes acquire more links ('rich-get-richer') than those that are less connected in a network, leading to the emergence of a few highly connected nodes which is referred to as hubs (*16*, *4*) or highly prominent nodes which play a vital role in shaping up the network. The resulting degree distribution of the network follows the power law, described in Equation 1. This power law distribution topology is radically different from the long established random graph theory of Erdős and Réyni (*17*), which assumed that complex systems are randomly created networks and eventually, shaped the researches in the area of complexity. However, a fundamental question always remained in the mind of complexity researchers: Is the real world network truly random in nature? The power law distribution described in Equation 1 is said to be a "stunning departure from Poisson distribution predicted by random network theory" (*4*).

In recent years, we have observed new advances in the area of network analysis which have demonstrated the scale free network behavior in many large scale real world networks including: Telephone call network (*18*); Worldwide Web (*6*); Internet (*19*);

metabolic reaction networks (*5*); software architecture (*20*); e-mail communication network (*21*); and distributed product development network in organization (*22*).

One of the important and primary uses of graph theory and network analysis is the identification of the most important actor(s) within a social network. Prominent actors are described as extensively involved in relationships with others (*23*, *24*). Hence, degree centrality has been used to describe the prominence of an actor in our email communication network. Number of emails sent by the employees to the actors within their respective communication networks is regarded as the degree centrality measure.

In our first experiment, we analysed the correlation of all actors' out-degree centrality values between two consecutive days. With few exceptions, there is a strong correlation for the actors' out-degree values between two consecutive days. The high correlation coefficient value implies that a high out-degree value for an actor on a particular day makes the same actor highly probable to have high out-degree values in the next consecutive day and vice-versa.

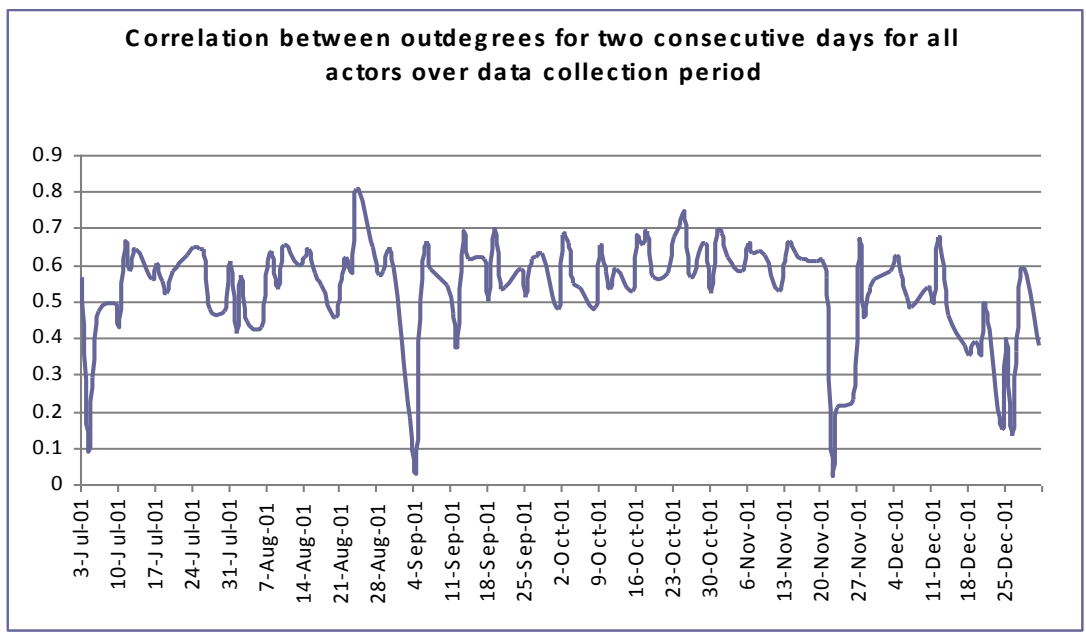

*Figure 1: Correlation coefficient values for the out-degrees of two consecutive days over the data collection period of 131 days.*

Second, each of these daily networks shows a distribution for out-degree centrality scores well described by power-law. Thus, a small number of highly connected nodes have greater importance in the connectivity of the entire communication network. Moreover, we found that there is a repetition of highly connected nodes in each daily network.

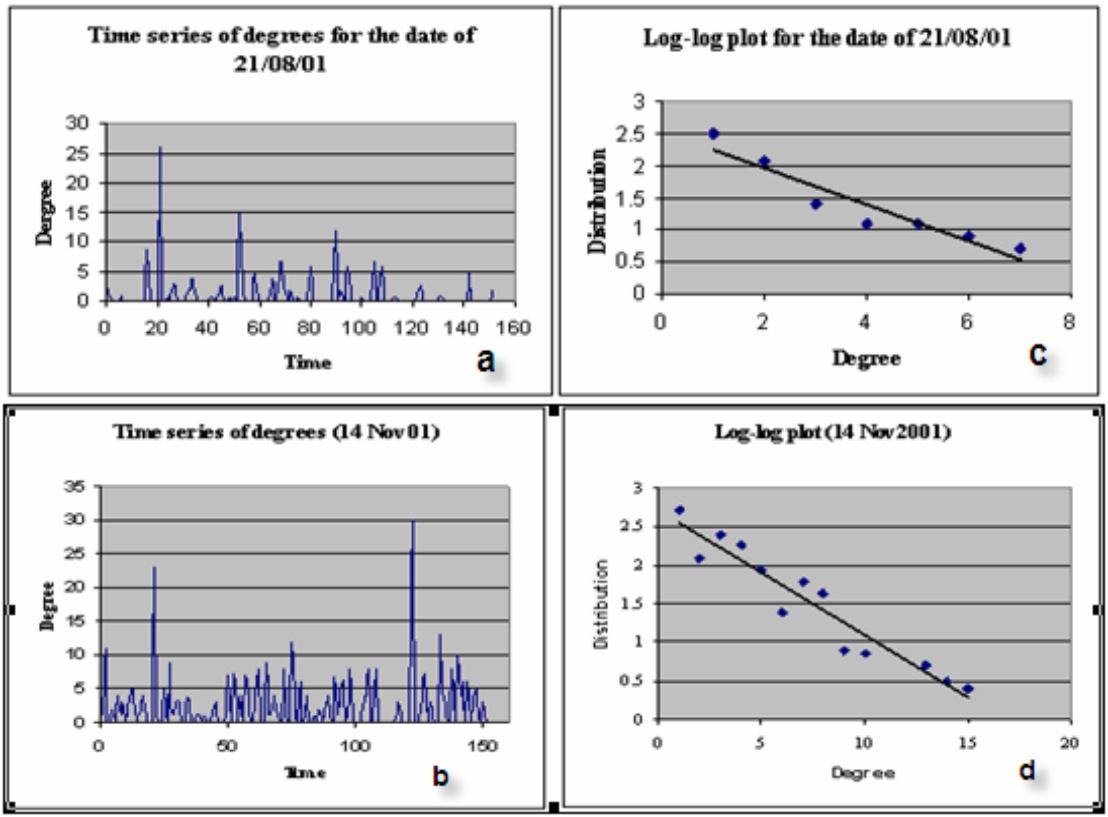

Figure 2: [**a-b**] Time series of degrees for randomly selected dates of 21 August 2001 and 14 November 2001. [**c-d**] are the corresponding log-log plot for the time series of degrees. The distribution of degree follow power-law distribution as they produce straight lines in the log-log plot.

After analysing the daily network, we measure the out-degree centrality scores for each of the identified prominent actor from everyday network, over the duration of the observation period. We found that most of the prominent actors exhibit stable time series.

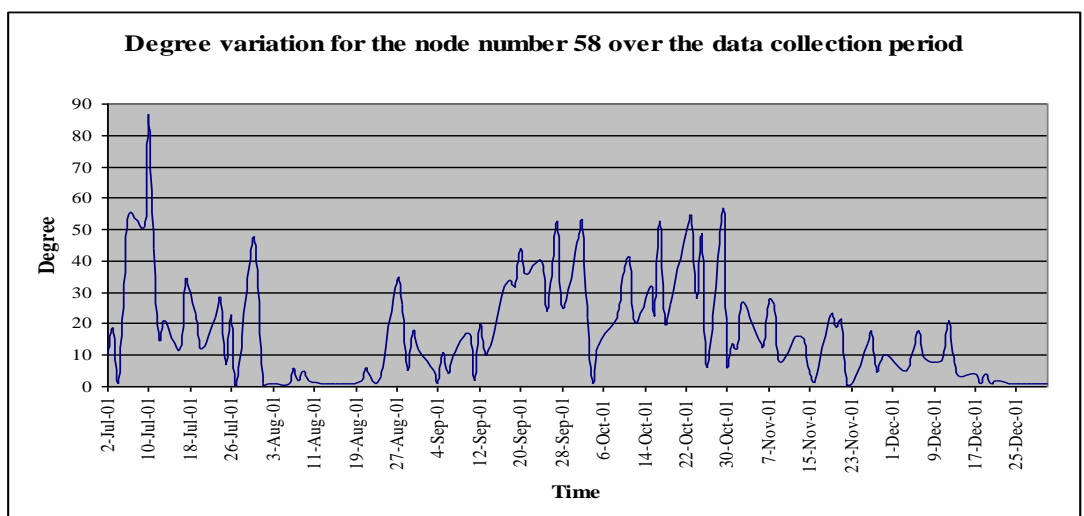

*Figure 3: Example of degree variation for the local hub (Node 58). Distribution of Node 58 does not follow the power-law theory.* This node is found most of the times (85 times) in the top-ten-rank list. Degree values are high in general with having low

values in few occurrences. Distribution for Node 58 does not follow the power-law distribution.

However, we also found that a few nodes exhibit a highly fluctuating time series. One such node is Node 12. Figure 4[**a**] demonstrates the fluctuating time series graph of the degree variations of Node 12.

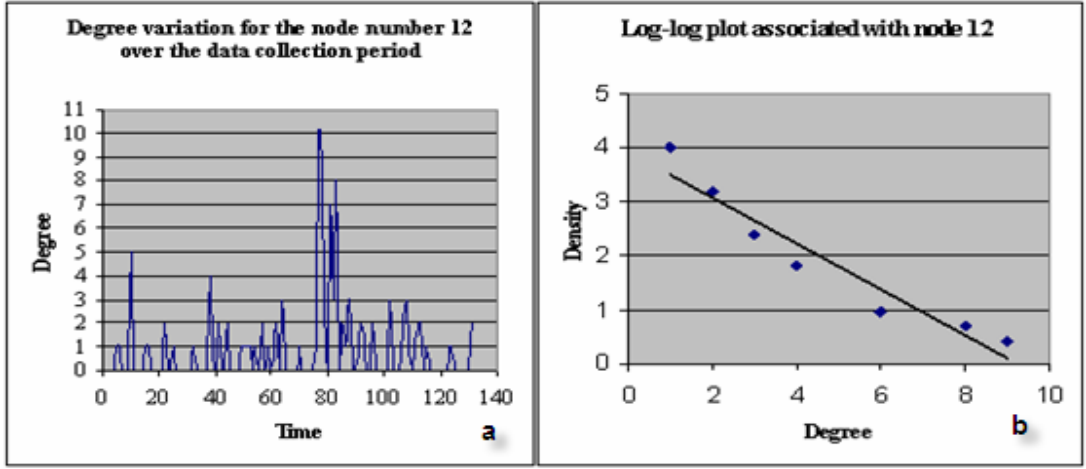

*Figure 4: [a] Example of fluctuating time series associated with the local hub (ID 12).* This node was found once acting as a prominent hub within the experiment period of 131 days of year 2001. *[b] The log-log plot for the degree distribution of this node also follows the power-law theory.* The log-log plot for the degree distribution of this node also follows the power-law distribution.

These results imply that some of the actors became very prominent in the network over time while some others became isolates. By summarising the results of figure 2, 3 and 4 we can conclude that the daily communication network of Enron employees (included in the dataset) followed the power-law distribution. However, when we look at the network structure of individual prominent actor, we find that some of them follow power-law distribution whereas some do not. We also identified top-ranking list of all actors for each day. We then determined for each pair of daily networks the percentage of nodes that appear in both ranking-lists. We found that the number and percentage of the out-degree centrality overlaps between any two daily networks increase with the increase of the size of the top-ranking list. This indicates same actors are repeatedly having the high centrality positions in the top ranking-list. Further, we compared the positions (ranks) of the most prominent nodes in daily networks with their positions in aggregated networks. We did not find any significant deviation between them. Out of top 10 actors who are frequently located in the top-ranking list of size ten of daily networks, 9 have also emerged as most prominent in the top-ranking list of aggregated networks. This implies that nodes are showing similar characteristics both in the daily and aggregated networks. Highly connected nodes in the daily networks have a similar role in the aggregated network. This result further confirms the scale free characteristic of the Enron communication network. Finally, we checked the percentage of degree values shown by top ten actors in the rank-list and compared it with the overall degree value shown by all actors in the network over the data collection period of 131 days. This result clearly demonstrates that only about 7% (top 10 nodes compared to 151) of the nodes exhibit around 60% of the total network degree values. Only a handful of actors became very prominent in the network. A few actors who are repeatedly located in the top ten list also showed high degree values in the daily networks as well as aggregated networks. These actors have become the prominent actors in the network over time, which further reinforces the 'rich-get-richer' phenomena observed in the scale free emerging communication network of Enron.

In this experiment, we studied both static and dynamic topology of social network analysis during Enron's disintegration period that captured the dynamics of communication networks. Both types of topologies exhibit similar characteristics of prominent actors reinforcing the power law distribution nature of scale free network. We did not find any significant fluctuation between the actor prominence in daily and aggregated networks of Enron. Our research showed similar outcomes with the works of Abello et al., (*18*); Albert et al. (*6*); Faulstos et al. (*19*); Jeong et al. (*5*); Valverde et al. (*20*); and Barabási (*16*).

Although Barabási (*16*) predicted a decade ago that a scale variant state of a network could be a generic property of many real life complex networks, it was not until recently when researchers (4) noted one of the most surprising discoveries of modern network theory, the 'universality' of the network topology. It has been noted that many real life networks, from the biological cell to the Internet and organizations, independent of their age, function and scope, display similar architectures. As a result of these similarities researchers from different disciplines embraced network theory as a common paradigm. However, it should also be noted here that this claim of 'universality' is not really universal. Some researchers (*25*) found that scale-free property is common in real life networks but not universal. For example, the co-authorship network of scientists shows power law distribution but with an exponential cutoff (*26*); the power grid network of western United States distribution is exponential (*27*); and for the social network of Mormons in Utah the distribution is Gaussian (*27*).

One of the many questions coming to our mind during this research is that - is there any functional advantage of scale free network topology? Albert et al. (*28*) found that there are practical advantages and disadvantages to it. They found that this type of network displays high degree of tolerance against random failures as only a few prominent hubs dominate their topology. However, the flip side is that such networks are extremely vulnerable to the attack on their hub(s). It has also been confirmed numerically and analytically by examining how the average path length and size of the prominent hubs depend on the number and degree of the nodes removed (*25*). It might well be true in the case of Enron's disintegration. We know that several of Enron's prominent hubs, including 2 former CEOs, a Chief Financial Officer, a number of vice presidents, and some senior management staff were involved (and subsequently implicated by the court) in the defrauding process of the organization. So, it is the local prominent hubs that exposed themselves to network vulnerability and eventually disintegrated Enron.

This study builds on an emerging stream of structural research that applies social network analysis to organizational email communication data in order to research important questions on organizational communication networks. With the increasing popularity of electronic communications, more widespread availability of such corpora, the increasing popularity of social network analysis and the growing sophistication of SNA tools, it is to be expected we can develop deeper insights into a wide range of organizational phenomena.